\documentclass[conference]{IEEEtran}

\usepackage{cite}      

\ifx\pdfoutput\undefined
\usepackage{graphicx}
\else
\usepackage[pdftex]{graphicx}
\fi

\usepackage{amsmath}   
\ifx\pdfoutput\undefined
\usepackage[hypertex]{hyperref}
\else
\usepackage[pdftex,hypertexnames=false]{hyperref}
\fi
\begin{document}

\title{Cosmic ray anisotropy studies with the Stockholm Educational Air Shower array}



%
\author{\authorblockN{Petter Hofverberg\authorrefmark{1} and
Mark Pearce\authorrefmark{1}}
\authorblockA{\authorrefmark{1}School of Engineering Sciences, Department of Physics\\
Royal Institute of Technology (KTH),
Stockholm, Sweden\\ Email: petter@particle.kth.se}}


\maketitle

\begin{abstract}
The Stockholm Educational Air Shower Array (SEASA) project has established a network of GPS time-synchronised scintillator detector stations at high-schools in the Stockholm region. The primary aim of this project is outreach. A part of the network comprises a dense cluster of detector stations located at AlbaNova University Centre. This cluster is being used to study the cosmic ray anisotropy around the knee. Each station consists of three scintillator detectors in a triangular geometry which allows multiple timing measurements as the shower front sweeps over the station. The timing resolution of the system has been determined and the angular resolution has been studied using Monte Carlo simulations and is compared to data. The potential of this system to study small and large scale cosmic ray anisotropies is discussed.
\end{abstract}


%
\IEEEpeerreviewmaketitle

\section{Introduction}
\begin{figure}[th]
\centering
\includegraphics[width=3.5in]{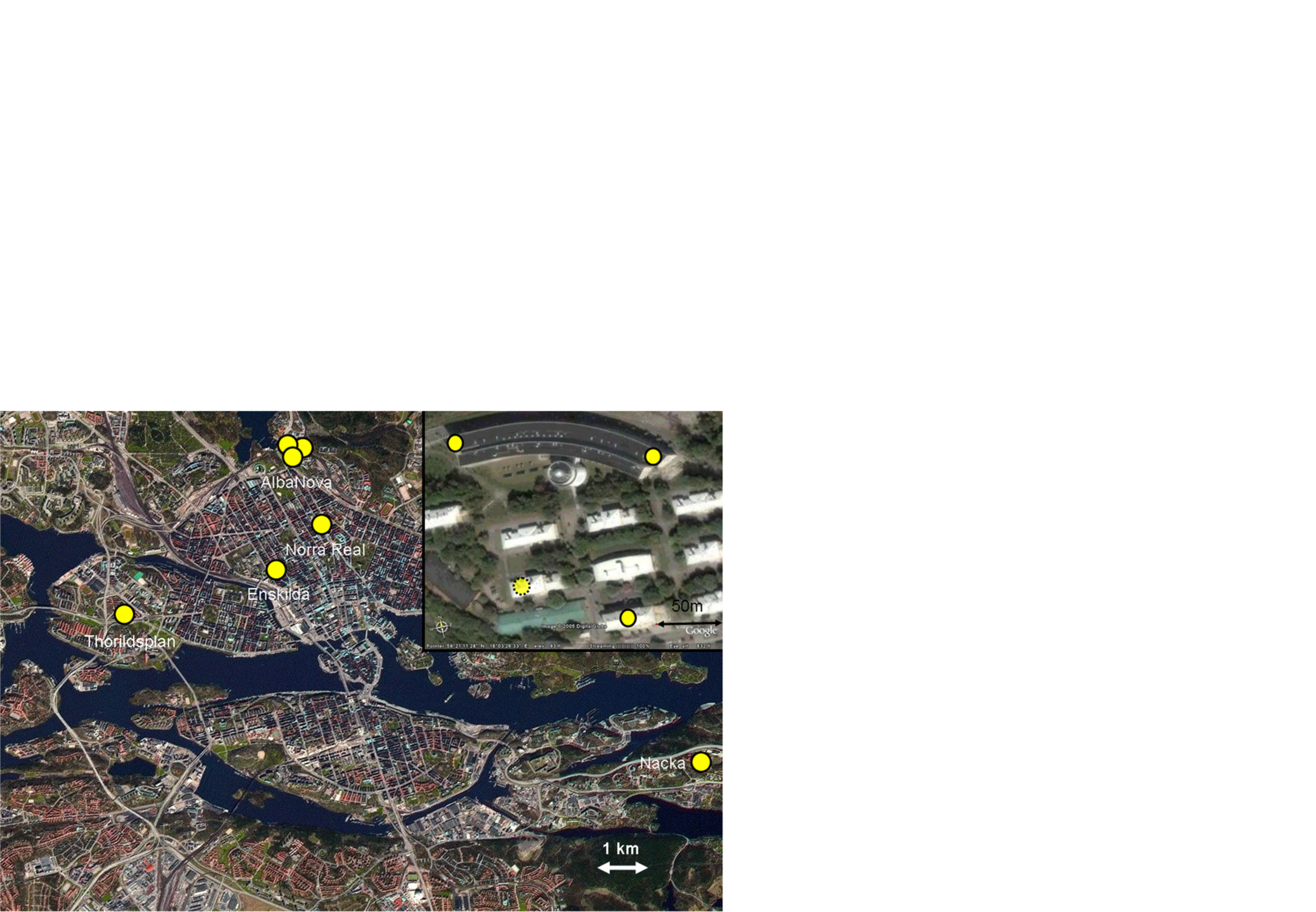} 
\caption{The seven stations constituting the SEASA air shower array. Four stations are located at high schools and are labeled after the corresponding schools. The three stations at the AlbaNova area are shown in greater detail in the inset picture in the upper right corner. The dotted station in this picture is a fourth station planned for installation.}
\label{pic:array}
\end{figure}

The 'Stockholm Educational Air Shower Array' (SEASA)~\cite{seasa}  project has established seven detector stations in the Stockholm area, distributed according to Fig.~\ref{pic:array}. Each detector station consists of three scintillator detectors, separated by approximately 15~m, which are read out by large area photomultipliers placed directly on top of the scintillators. The design of the detectors is explained in detail in~\cite{mylic}. Four of the stations have been built by high school students and are located on the attics or roofs at their high schools. The separation between these stations are in the order of kilometers and the energy threshold to trigger multiple stations is therefore above $10^{18}$~eV. To be able to study lower energetic cosmic rays a more dense cluster of three detector stations has been deployed at the AlbaNova University Area (the \emph{AlbaNova array}). This cluster detects air showers with energies above $~10^{16}$~eV and is being used to study the cosmic ray arrival distribution in this energy regime.

To study the cosmic ray anisotropy the direction of the primary cosmic ray must be possible to reconstruct. This can be done as the air shower front travels in essentially the same direction as the primary particle. Assuming a flat shower front, the direction can be determined by measuring the time difference between hit detector stations. By fitting the geometry of the detectors and the trigger times to the shower plane the incident angles of the shower can be reconstructed. With a fixed geometry the accuracy of the reconstruction  ultimately depends on the trigger time resolution, set by the GPS system. The time resolution of the GPS systems is therefore investigated in section~\ref{sec:timeres}.

The pointing accuracy, or angular resolution, of the AlbaNova array has been assessed by simulations and this is described in section~\ref{sec:angres}. The dependence of the angular resolution on the timing accuracy is also investigated here, as well as the trigger efficiency of the array. The final section~\ref{sec:anisotropy} describes the methods used to study cosmic ray anisotropies. Finally, the hypothesis of a uniform flux of cosmic rays are tested using data taken during approximately six months of operation of the AlbaNova array.

\section{Time resolution of the system}
\label{sec:timeres}

To test the performance of a GPS system, the offset between the
time-tags produced by the GPS system subject for measurement and a reference system, fed with a common
trigger signal, is investigated.

The output from the GPS card is a 1 Pulse Per Second (PPS) signal with an accuracy of
$\pm$ 25~ns. The PPS can only be emitted on the rising edge of the
GPS-cards internal 100~MHz oscillator, which introduces a built in
uncertainty. The output from the GPS card also contains a negative
sawtooth correction. This correction is a prediction of how early or late
the next PPS signal will be due to the limitations of the internal
100~MHz oscillator. With the aid of this correction the PPS should
be accurate to within 5~ns according to the developer of the GPS
cards~\cite{gps}.

The principle of the test is as follows. For every trigger, a time
stamp from each GPS card is retrieved. These time stamps should be
identical in a perfect system. The time stamp is provided by the
sawtooth corrected PPS and a 100~MHz oscillator implemented in
the Programmable Logic Array (PLA) in the readout electronics. The information in the PPS signal gives the time within the second and the oscillator determines the trigger time relative to the PPS with a 5~ns resolution. A self calibrating system for the 100~MHz crystals were
used to compensate for differences in the crystal frequencies and
variations in the crystal frequency. To be self calibrating, the
system counts the number of oscillations between PPS's, and uses
this value to calibrate the number of oscillations from the
trigger to the PPS. All measurements in the test were done with a
satellite mask angle of ten degrees to exclude unreliable time
measurements from satellites close to the horizon.

\begin{figure}[bh!]
\centering
\includegraphics[width=3.5in,angle=-90]{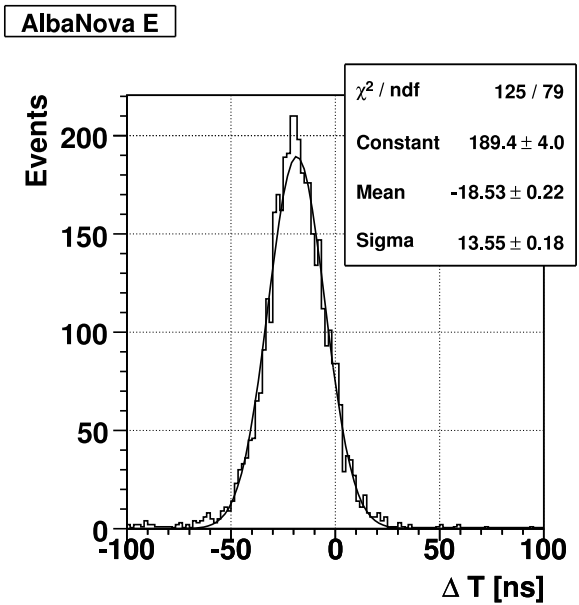}
\caption{The differences between time stamps from AlbaNova E and AlbaNova W (the reference system) when fed with a common trigger signal.}
\label{gpstest}
\end{figure}

The result of the measurement for the AlbaNova E station is shown in Fig.~\ref{gpstest}. The offset of -18.5~ns for the mean value changes sign when the GPS cards are exchanged, indicating a systematic error between these two. This effect can be canceled by calibrating each card against a ``standard card'', and then correcting the time stamps from each card accordingly. The standard deviation of 13.6~ns is the time resolution for this pair of GPS systems. The time resolution of a single GPS system is 9.6~ns considering the GPS systems equal and independent~\cite{mylic}. The time resolution for each of the seven GPS systems in the present array were measured to be less than 15~ns. 

In the test, the GPS cards mostly tracked the same satellites. It
is inevitable that detector stations spread out over a larger area
will have different sets of satellites visible to them. A test was
therefore conducted where the GPS cards were configured to use
independent sets of satellites. The standard deviation then increased by approximately 50~\%.

\section{Simulation of the angular resolution}
\label{sec:angres}
The angular resolution of the sub-array at AlbaNova has been assessed with Monte Carlo simulations, using the simulation engine AIRES. Primary particles with energies above $10^{16}$~eV following the power-law $\frac{dE}{dN}\sim E^{-3.0}$ were generated and injected at the top of the atmosphere. The lower energy was chosen considering the energy threshold of the AlbaNova array, known to be slightly higher than this value from simulations (see section~\ref{sec:triggeff}). The injection angle was sampled from a uniform distribution. 
A total of 2000 cosmic rays with the above properties were generated and the ground particles from each shower were repeatedly used to hit the ground at different offsets relative to the detector array. The impact coordinates of the shower core was set to follow a $9\times9$ grid with a node separation of 50~m and the origin placed in the centre of the detector array. The number of detected showers outside this area is negligible and can thus be disregarded. 
A detector is triggered if it is hit by an electron, muon or heavier charged particle. If a photon hits the detector it is triggered with a 1\% probability. This value corresponds to the probability that a photon deposit at least 1~MeV in a 1~cm thick scintillator. To simulate imperfections in the GPS time tagging a time jitter $\sigma_t$ sampled from a Gaussian with a standard deviation of 15~ns is added to the time of the hit. This value is based on the time resolution measurements presented in section~\ref{sec:timeres}.

\subsection{The trigger efficiency}
\label{sec:triggeff}
\begin{figure}[bh!]
\centering
\includegraphics[width=3.5in,angle=-90]{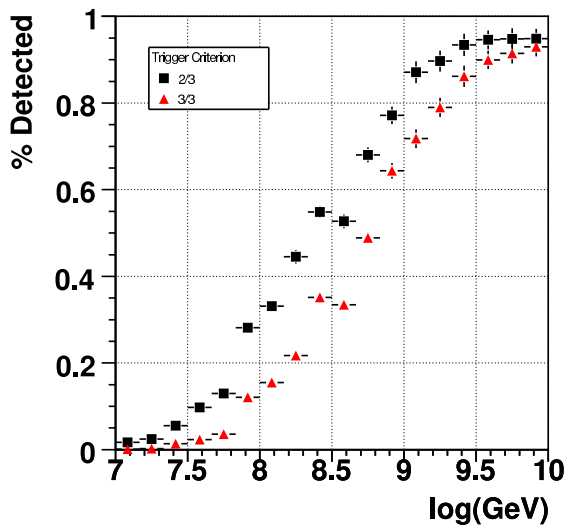}
\caption{The trigger efficiency for the AlbaNova sub-array for two different trigger criteria.}
\label{pic:triggeffsuper}
\end{figure}

The trigger efficiency of the air shower array has been determined from the simulations simply by dividing the number of detected showers with the number of incident cosmic rays for bins of energy. Fig.~\ref{pic:triggeffsuper} shows the result of the simulation for the two different trigger criteria: $3/3$ detectors hit or at least $2/3$ detectors hit. The energy threshold is consequently around $10^{16.5}$~eV and the most probable energy slightly higher, $\sim10^{17}$~eV. Thus, SEASA detects cosmic rays with energies \emph{below} the knee with this setup. The trigger efficiency is slightly underestimated in this study as the simulation program does not propagate particles close to the core and a detector station is thus not triggered if an air shower strikes directly on top of it. The underestimation is believed to be around 5~\% as the efficiency should saturate at 100~\% for large energy showers (see Fig.~\ref{pic:triggeffsuper}).. The underestimation is believed to be equal for all energies. 

\subsection{The angular resolution of the AlbaNova array}
The angular resolution of the detector array is derived by comparing the shower direction reconstructed from the timing information of the hit detectors with the direction of the primary particle inputted in the simulation. The precision of the reconstruction is measured as the angular distance between the two shower directions, characterised by the parameters $(\theta_1,\phi_1)$ and $(\theta_2,\phi_2)$, where $\theta$ is the zenith angle and $\phi$ the azimuthal. The angular distance between two directions is then calculated as
\begin{equation}
\Psi = cos^{-1}\left(cos\theta_1cos\theta_2+sin\theta_1sin\theta_2cos(\phi_1-\phi_2)\right)
\end{equation}

\begin{figure}[th]
\centering
\includegraphics[width=3.5in,angle=-90]{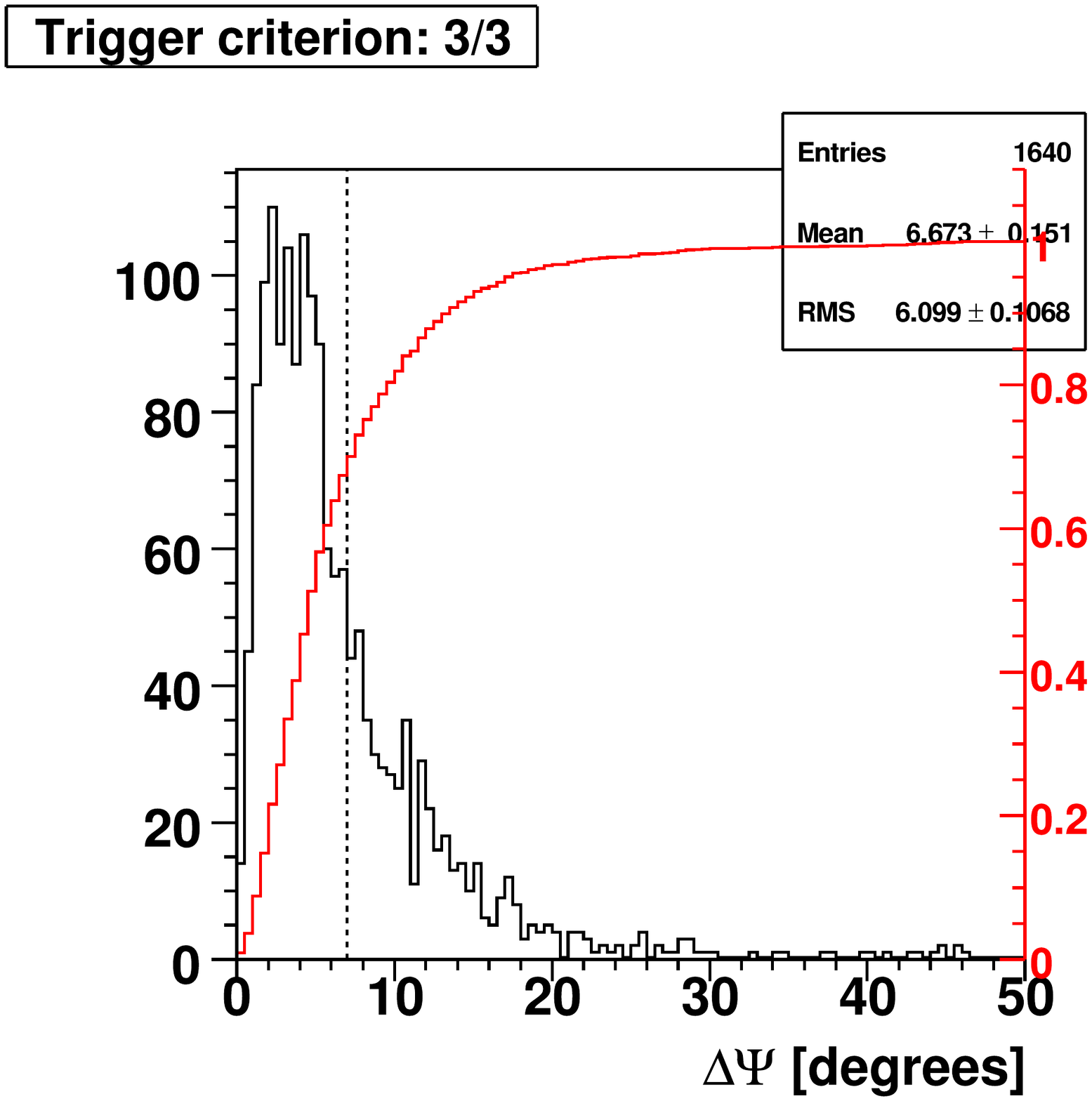}
\caption{The histogram shows the angular distance between the reconstructed angle and the true angle of the primary particle for the $3/3$ trigger criteria. The axis to the right in the plot corresponds to the integral of the angular distance, superimposed on the histogram. The dotted vertical line marks the 68~\% level of the integral and is the definition of the accuracy of the angle reconstruction for the three station setup for SEASA.}
\label{pic:angleres}
\end{figure}

The angular resolution of the detector array is defined as the angular distance which contains $68\%$ of the reconstructed angles. This is the most common way to define the angular resolution and therefore makes it straight-forward to compare the result from SEASA to other air shower arrays. Some experiments use the $50\%$ level as the angular resolution and this is therefore included in the results below. The distribution of the angular distance is plotted in Fig.~\ref{pic:angleres} below for the $3/3$ trigger criteria. Superimposed on the distribution is the integral of the histogram with the corresponding axis to the right in the plot. The vertical dotted line marks the $68\%$ level of the integral and thus the angular resolution. 

The angular accuracy of the AlbaNova array is summarised in Table~\ref{tab:angres} for the two trigger criteria. The errors have been calculated by randomly regenerating the histogram of the angular distance a large number of times from the true distribution. The RMS of the derived distribution of the angular resolution is then used as the error for the true angular resolution.  The angular accuracy, as shown in Table~\ref{tab:angres}, is the same for both trigger modes, within statistical fluctuations. This is an important conclusion and makes it possible to increase the trigger rate by loosening the trigger criterion without compromising the angular accuracy of the array. 

\begin{table}[h]
\centering
\begin{tabular}{l|l|l}
Trigg. Crit.  & Resolution $68\%$ & Resolution $50\%$ \\
\hline
$\Psi_{2/3}$ & {\bf 6.5$\pm$ 0.3} & 4.5$\pm$ 0.2\\
$\Psi_{3/3}$ & {\bf 7.0$\pm$ 0.3} & 4.5$\pm$ 0.2\\ 
\end{tabular}
\caption{The angular resolution of the detector array for two different trigger criteria and levels. The $68\%$ level is used for the definition of the angular resolution for SEASA.}
\label{tab:angres}
\end{table} 

The time resolution of 15~ns used in the simulations is only valid for the AlbaNova array, mainly due to two reasons: First, the relative time accuracy between two GPS setups decreases with separation distance because of differences in the atmosphere along the satellite-antenna path lengths. Secondly, the set of visible GPS satellites can change when the separation between the antennas is large, which has a negative influence on the relative time resolution. The effect of the time resolution on the angular resolution has been investigated by varying the Gaussian time jitter $\sigma_t$, introduced in the last section, and the result is presented below in Fig.~\ref{pic:angresvstime}. The angular accuracy is seen to decrease approximately linearly with the time resolution.

\begin{figure}[h]
\centering
\includegraphics[angle=-90,width=4.25in]{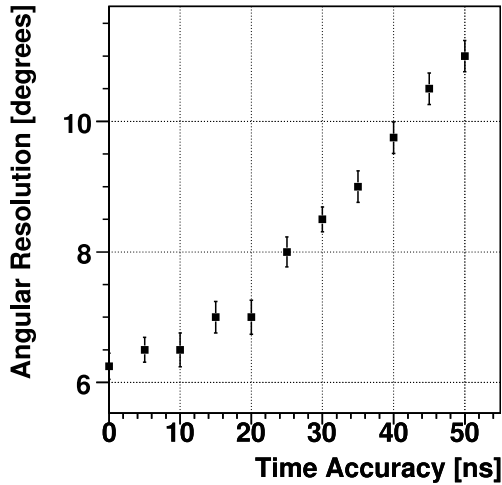}
\caption[Angular resolution vs time resolution.]{The angular resolution as a function of the time resolution in the system. For the array at AlbaNova, a time resolution of 15~ns is used, but this is expected to increase with a larger array as described in the text.}
\label{pic:angresvstime}
\end{figure}

\subsection{Validation of the simulations}
To confirm the validity of the simulations the difference between the reconstructed angles by the AlbaNova array and by each station in the array are derived for real and simulated data. These are compared below in Fig.~\ref{pic:comparisonsimreal} and the agreement is relatively good for AlbaNova W and E. The shift in the histograms between simulated and real data for the azimuthal distributions are likely caused by the crude measurement of the local coordinate system for the detectors in each station. The agreement is however poor for AlbaNova S. This is most likely due to the effect of the roof and walls that surround this station. Simulations that takes this into account will be performed in the future. However, the results in Fig.~\ref{pic:comparisonsimreal} are a good indicator that the simulations are correct. The difference between the reconstructed angles are in fact slightly smaller for real data indicating that the performance of the array may be better than the simulations show. 

\begin{figure}[h]
\centering
\includegraphics[angle=-90,width=4.5in]{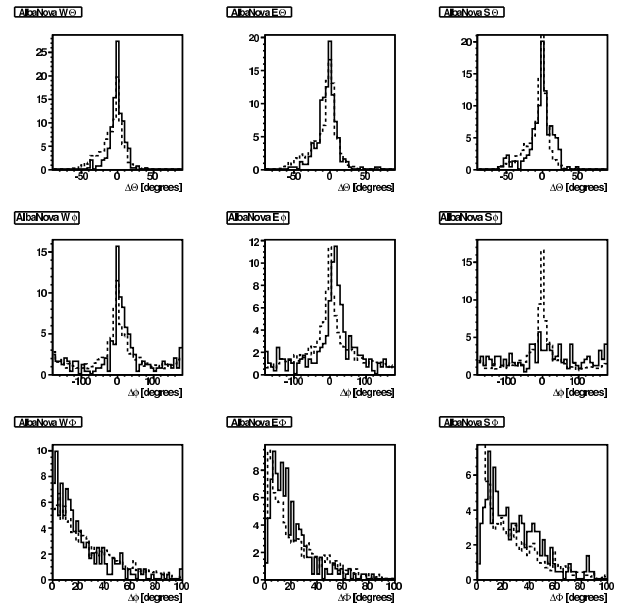}
\caption{A comparison between real data (filled) and simulations (dotted) of the difference between the reconstructed angles by the entire AlbaNova array (super angles) and reconstructed angles by single detector stations (station angles). The top row shows the difference in reconstructed zenith angle $\Theta$ for the three detector stations. The middle row shows the azimuthal angle $\phi$ and the bottom row the space angle difference $\Phi$. The agreement between data and simulations are good except for the azimuthal distribution for AlbaNova S which most likely is caused by the fact that this station is located inside a building. Notice that the difference between super angles and station angles are somewhat smaller for real data than for the simulations.}
\label{pic:comparisonsimreal}
\end{figure}

\section{Anisotropy searches}
\label{sec:anisotropy}
This section presents a preliminary study of the methods that can be
used by the SEASA experiment to search for small and large scale
cosmic ray anisotropies. To date, the collected statistics are poor due to the
short period of data taking, the small exposure and of the numerous tests that have been
performed during the initial phase of the project. SEASA aims to
lower the energy threshold in the future, by adding more stations and
loosening the trigger criteria, thereby increasing the rate of
detected showers. More accurate studies of small and large scale
anisotropies will then be feasible.

The search for anisotropies relies heavily on the
estimation of the number of cosmic rays expected from each direction
in the sky assuming a uniform flux over the celestial sphere. Such a estimation is henceforth called a \emph{coverage
  map}. An unbiased coverage map is crucial in order to separate true anisotropies from acceptance effects. This is relatively straight-forward for ultra high energy cosmic rays ($E>10^{18}$~eV) where the total acceptance almost exclusively depends on the geometrical acceptance of the experiment. The derivation of the coverage map is more complicated at lower energies as variations in the atmospheric conditions then influences the detector acceptance heavily. This is balanced somewhat by the large number of low energy events.

\subsection{The coverage map}
\label{sec:coverage}
\begin{figure}[h]
\centering
\includegraphics[width=3.5in]{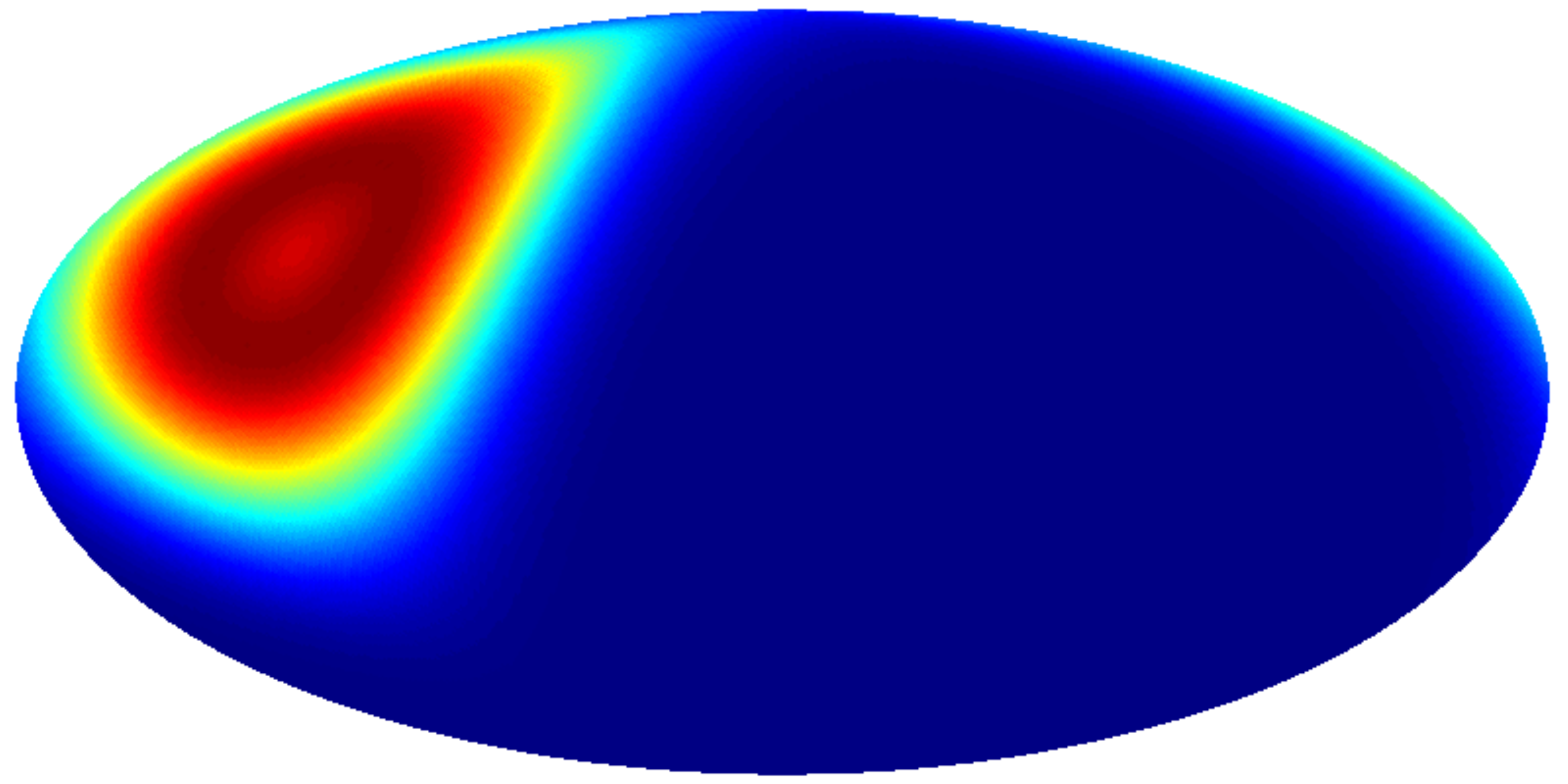}
\caption{The coverage map in galactic coordinates.}
\label{pic:coveragemap}
\end{figure}

The coverage map for a given data set is obtained by integrating the acceptance of the experiment over the data taking period. The acceptance generally depends on weather conditions and the direction in the sky, characterised by declination and right ascension. This corresponds to ($\theta(t)$,$\phi(t)$) at UTC~\footnote{Coordinated Universal Time} t, in horizontal coordinates. Using the simplification that the acceptance is independent of azimuthal angle $\phi$ and weather conditions, the acceptance is a function of zenith angle $\theta$ only. The zenith angle distribution has been shown~\cite{auger_coverage} to be almost unaffected by anisotropies, and this distribution is therefore used as a basis when calculating the acceptance. The function
\begin{equation}
P(\theta)=Acos(\theta)sin(\theta)\frac{1}{1+exp\left(\frac{\theta-\theta_0}{\Delta\theta}\right)}
\end{equation}
is fitted to the measured zenith angle distribution and converted to declination acceptance through the formula $a(\theta)=\frac{1}{sin(\theta)*P(\theta)}$ (solid angle effect). A coverage map, that only depends on declination, is then generated by integrating the acceptance over one sidereal day.
\begin{equation}
\label{eqn:oneday}
W(\delta)=\int_{0}^{24h}a(\theta)dt
\end{equation}
 The resultant coverage map in galactic coordinates can be seen in Fig.~\ref{pic:coveragemap}

\subsection{A first measurement of the cosmic ray anisotropy}
\label{sec:measurement}
The hypothesis that the flux of cosmic rays is isotropic can be tested using data from the SEASA experiment. A simple approach is to derive the angular two point correlation function $w(\Phi)$. In its angular form it is defined by the expression 
\begin{equation}
\label{eqn:twopoint}
\delta P = N[1+w(\Phi)]\delta\Omega
\end{equation}
 where $\delta P$ is the probability to find a second object within an angular distance of $\Phi$ from the primary object within an area $\delta\Omega$ if the mean object density is N. The two point correlation function thus represents an ``excess probability'' above what is expected from an isotropic distribution.

\begin{figure}[h]
\centering
\includegraphics[width=3.5in]{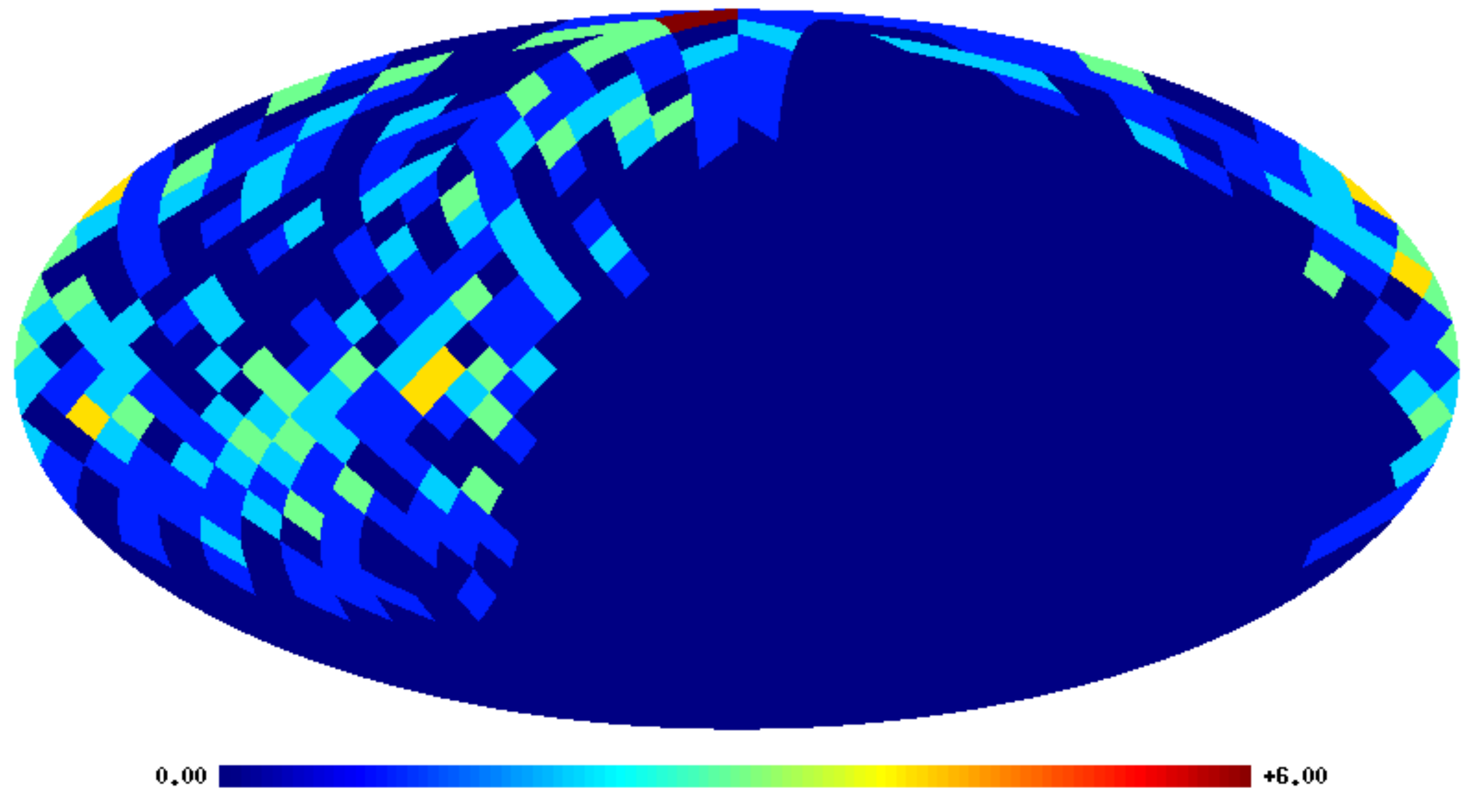}
\caption{A sky map in galactic coordinates showing the detected events.}
\label{pic:eventmap}
\end{figure}

\begin{figure}[h]
\centering
\includegraphics[angle=-90,width=4.0in]{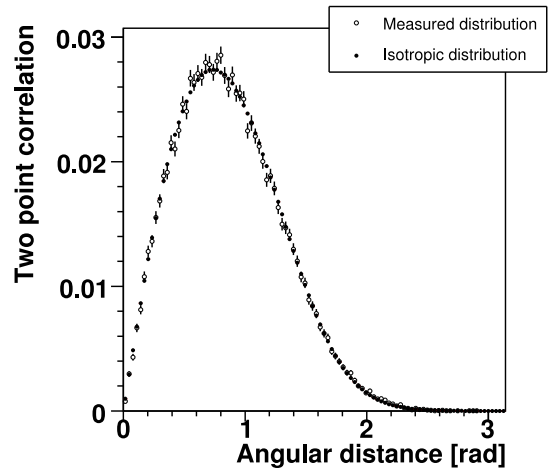}
\caption{The distributions of the two point distance for measured events and randomly generated isotropic events.}
\label{pic:corr}
\end{figure}

The measured sky-plot of cosmic rays is plotted in galactic
coordinates in Fig.~\ref{pic:eventmap}. The two-point correlation distribution is derived
by calculating the distance between all possible pair of events for
this data set. To compare this to the hypothesis that the arrival
distribution is isotropic a second two-point correlation distribution
is derived from a randomly generated isotropic distribution convoluted
with the coverage map derived in the previous section. Possible
deviations of the first distribution from the second then reveals
anisotropies of the cosmic ray arrival distribution. Both correlation
distributions mentioned above are plotted in Fig.~\ref{pic:corr}. The probability that the observed flux is a random sampling from an isotropic flux is checked with a Kolmogorov test and it is found to be 82\%. The hypothesis of an isotropic flux is therefore supported.

\section{Conclusions}
A measurement of the cosmic ray anisotropy has been made using the AlbaNova array which forms a sub array of the SEASA outreach project. The result favors a scenario with a uniform flux of cosmic rays in the energy regime above $10^{16}$~eV, and therefore agree with previous measurements, for example KASCADE~\cite{kascade}. It is a crude measurement of the cosmic ray anisotropy but it shows that the array is stable and in particular that the GPS timing is reliable. The measurement will be improved in the near future by adding a fourth station at the AlbaNova area which increases the exposure of the array and also improves the shower angle reconstruction. A future improvement is also to enhance the read out electronics with a pulse sampling device. This would make it possible to determine the shower core and curvature and thereby improving the angular resolution of the array. This would allow anisotropy searches on smaller scales than what has been possible up to now.




%





\end{document}